\documentclass[twocolumn,showpacs,preprintnumbers,amsmath,amssymb]{revtex4}
\def\be{\begin{equation}}
\def\ee{\end{equation}}
\def\bea{\begin{eqnarray}}
\def\eea{\end{eqnarray}}
\def\dT{{\delta{\bf T}_1}}
\def\dTT{{\delta{\bf T}_2}}
\def\drho{{\delta\rho_1}}
\def\drhorho{{\delta\rho_2}}
\def\dP{{\delta P_1}}
\def\dPP{{\delta P_2}}

\def\L{{\pounds}}
\def\H{{\cal H}}

\def\R2{{{\cal{R}}_2}}
\begin{document}
%
\title{Evolution of second-order cosmological perturbations}
\author{Karim A.~Malik$^1$ and David Wands$^2$}
%
\affiliation{ 
$^1$GRECO, Institut d'Astrophysique de Paris, C.N.R.S.,
98bis Boulevard Arago, \\
75014 Paris, France\\
$^2$Institute of Cosmology and Gravitation, University of Portsmouth,
Portsmouth~PO1~2EG, United Kingdom}
\date{\today}
\begin{abstract}
  We present a method for constructing gauge-invariant cosmological
  perturbations which are gauge-invariant up to second order. As an
  example we give the gauge-invariant definition of the second order
  curvature perturbation on uniform density hypersurfaces. Using only
  the energy conservation equation we show that this curvature
  perturbation is conserved at second order on large scales for
  adiabatic perturbations.
\end{abstract}


\pacs{98.80.-k \hfill Class.\ Quant.\ Grav.\ {\bf 21}, L65 (2004),
astro-ph/0307055v3}

\maketitle 

\section{Introduction}

In this letter we report new results for the study of second-order
perturbations about a Friedmann-Robertson-Walker (FRW) spacetime. The
study of second-order perturbations to date has been limited by two
main problems: firstly the difficulty in defining truly
gauge-invariant perturbations at second order, and secondly the
complexity of the resulting Einstein equations~\cite{Salopek,marco}.

We address the first of these challenges by extending an approach to
the construction of gauge-invariant variables previously advocated for
first-order perturbations~\cite{firstpaper}. We make an unambiguous
physical definition of the perturbation, and by building this into
the mathematical description of the physical perturbation, construct
a gauge-invariant quantity. As an example, we give a gauge-invariant
definition of the curvature perturbation on uniform density
hypersurfaces. 

We avoid much of the complexity of the second-order field equations by
considering only the local energy conservation
equation~\cite{WMLL,LW03} in the large-scale limit where we neglect
all spatial derivatives. This gives us a simple result establishing
the constancy of the large-scale curvature perturbation on uniform
density hypersurfaces, up to and including second order, for adiabatic
perturbations.

To use, for example, the observed Gaussian distribution of
cosmic microwave background (CMB) anisotropies~\cite{KomatsuWMAP} to
test models for the origin of structure in the very early universe,
there is an implicit assumption that there is negligible growth of
second order perturbations on large scales. 
Simple models of inflation produce an almost Gaussian distribution of
density perturbations~\cite{Gangui}, and recent studies have shown
that 
the comoving curvature perturbation does not evolve at second order on
large scales during slow-roll
inflation~\cite{Acquaviva,Maldacena,Rigopoulos}. But non-linear
evolution from the end of inflation up until the time of
last-scattering of the CMB could produce deviations from a Gaussian
distribution.
The existence of a constant curvature perturbation at second order
shows that the observed distribution of CMB anisotropies can be used
to directly constrain the distribution of perturbations produced in
the very early universe in simple inflation models that predict
adiabatic density perturbations after inflation.


\section{Second-order perturbations}


Observations on scales close to the Hubble scale seem to be consistent
with an almost homogeneous and isotropic spacetime that can be
described by small perturbations about a Friedmann-Robertson-Walker
metric.

Any tensorial quantities can be split into a
homogeneous background and inhomogeneous perturbation
\be
\label{defT}
{\bf T}(\eta,x^i) = {\bf T}_0(\eta) + \dT(\eta,x^i) + \frac12
\dTT(\eta,x^i) + \ldots
\ee
where we use subscripts $1$ and $2$ to denote first and second order
perturbations. 

We will consider perturbations about a spatially flat FRW
background metric
\be
ds^2 = a^2 \left[ -d\eta^2 + \delta_{ij} dx^i dx^j \right] \,.
\ee
where $\eta$ is conformal time and $a=a(\eta)$ is the scale factor.
The metric tensor including second-order perturbations can be written as
\bea
\label{metric1}
g_{00}&=&-a^2\left(1+2\phi_1+\phi_2\right) \,, \\
g_{0i}&=&a^2\left(B_{1i}+\frac{1}{2}B_{2i}\right) \,, \\
g_{ij}&=&a^2\left[\left(1-2\psi_1-\psi_2\right)\delta_{ij}
+2C_{1ij}+C_{2ij}\right]\,.
\eea
%
%
We will refer to $\psi$ as the curvature perturbation as it describes
the intrinsic scalar curvature of constant-$\eta$ hypersurfaces on
large scales.

Perturbations can be split into scalar, vector, and tensor modes,
according to their transformation behaviour on spatial 3-hypersurfaces
\cite{Bardeen}. For instance we can write
\be
2C_{ij} = 2E_{,ij} + F_{i,j} + F_{j,i} + h_{ij} \,,
\ee
where $E$ is a scalar perturbation, $F_i$ is a divergence-free vector,
and $h_{ij}$ a transverse, trace-free tensor perturbation.

%
%

\section{Gauge transformations}

Under a second-order coordinate transformation
\be
\label{coordtrans2}
\widetilde{x^\mu} = x^\mu+\xi_1^{\mu}
+\frac{1}{2}\left(\xi^{\mu}_{1,\nu}\xi_1^{~\nu}+ \xi_2^{\mu}
\right)   \,,
\ee
%
%
any tensor ${\bf T}$ and its perturbations defined in Eq.~(\ref{defT})
transform as \cite{marco}
\bea
\label{Ttrans}
\widetilde \dT
&=& \dT + \L_{\xi_1} {\bf T}_0  \nonumber\\
\widetilde \dTT
&=& \dTT +\L_{\xi_2} {\bf T}_0 +\L^2_{\xi_1}
{\bf T}_0 + 2\L_{\xi_1} \dT
 \,.
\eea

Thus under a first-order transformation
$\xi_1^\mu=(\alpha_1,\beta_1^i)$, a scalar quantity such as the density,
$\rho$, transforms at first-order as
\be
\label{defrho1}
\widetilde\drho = \drho + \rho_0'\alpha_1 \,,
\ee
while at second order, writing $\xi_2^\mu=(\alpha_2,\beta_2^i)$, we have
\bea
\label{defrho2}
\widetilde\drhorho &=& \drhorho 
+\rho_0'\alpha_2+\alpha_1\left[
\rho_0''\alpha_1+\rho_0'{\alpha_1}'+2\drho'\right]\nonumber\\
&& +\left(2\drho+\rho_0'{\alpha_1}\right)_{,i} \beta_1^i
\,.
\eea

For the first order curvature perturbation we have
\be
\label{defpsi_1}
\widetilde \psi_1 = \psi_1-\H\alpha_1 \,,
\ee
where $\H\equiv{a'}/{a}$, while at second order we get from 
Eq.~(\ref{Ttrans}),
\bea
\label{defpsi_2}
\widetilde \psi_2 &=& \psi_2
-\alpha_1\left[\H{\alpha_1}'
+\left(\H'+2\H^2 \right)\alpha_1
-2\psi_1'-4\H\psi_1\right]\nonumber \\
&&-\H\alpha_2 -\left(\H\alpha_1-2\psi_1\right)_{,i}\beta_1^i
 \,.
\eea
%

\section{Gauge choices and gauge-invariant variables}

A gauge-invariant theory of linear perturbations about FRW metric was
proposed by Bardeen \cite{Bardeen} and subsequently developed by many
authors (for example see Refs.\cite{KS,MFB,Durrer,Hwang,thesis}). No
such gauge-invariant formalism has been developed for non-linear
cosmological perturbations~\footnote
{After completing this work, our attention was drawn to a recent paper
  by Nakamura~\cite{Nakamura} where a general procedure is proposed
  for constructing gauge-invariant perturbations upto third order
  of a generic spacetime.}.
%
According to the Stewart-Walker lemma \cite{SW73} any truly
gauge-independent perturbation must be constant in the background
spacetime. This apparently limits ones ability to make a
gauge-invariant study of quantities that evolve in the background
spacetime, e.g., density perturbations in an expanding cosmology. 

In practice one can construct gauge-invariant definitions of
unambiguous, that is physically defined, perturbations. These are not
unique gauge-independent perturbations, but are gauge-invariant in the
sense commonly used by cosmologists to define a physical perturbation.
We draw a distinction here between quantities that are automatically
gauge-independent, i.e., those that have no gauge dependence (such as
perturbations about a constant scalar field), and quantities that are in
general gauge-dependent (such as the curvature perturbation) but can have a
gauge-invariant definition once their gauge-dependence is fixed (such as
the curvature perturbation on uniform-density hypersurfaces).     
Although this approach has been widely used, at least implicitly, to
construct gauge-invariant quantities at
first-order~\cite{firstpaper,thesis}, it has not previously been used
at higher-order. In this letter we show that it is possible to
define gauge-invariant quantities at second-order corresponding to
physical perturbations.




\subsubsection{Uniform density hypersurfaces}

The uniform density hypersurfaces are defined by setting
$\widetilde{\delta\rho}=0$ to 
the required order in perturbation theory. We
find that for a specific spatial gauge this leads to a specific
temporal gauge to 
the required order.

{}From Eq.~(\ref{defrho1}) we see that setting $\widetilde\drho=0$ to
first-order requires a gauge shift from an arbitrary gauge
\be
\label{xi0rho}
\widetilde\alpha_1 = - \frac{\drho}{\rho_0'} \,.
\ee
Leaving for the moment the spatial gauge dependence, 
we see from Eq.~(\ref{defrho2}), and using Eqs.~(\ref{xi0rho}) to fix
the first-order temporal gauge shift, that to second order we require
\be
\label{beta0rho}
\widetilde\alpha_2
=
-\frac{1}{\rho_0'}\left[
\drhorho-\frac{1}{\rho_0'}\drho'\drho
 + \drho_{,i} \widetilde\beta_1^i
\right]\,.
\ee

To completely fix the second-order temporal gauge shift
(\ref{beta0rho}) picking out uniform density hypersurfaces we must
also specify the first-order spatial gauge shift $\widetilde\beta_1^i$. 
For example, a natural choice is to pick worldlines comoving with the fluid. 
The fluid 3-velocity transforms as
\be
\label{defv}
\widetilde{v}^i = v^i-\beta^{i\prime} \,.
\ee
Thus from an arbitrary spatial gauge we can transform
to the comoving gauge by the spatial gauge transformation
\be
\label{betai}
\widetilde\beta^i = \int v^i d\eta \,.
\ee
In this case there is a constant of integration corresponding to
the arbitrary choice of spatial coordinates labelling the worldlines
on an initial time-slice.

We are now able to construct gauge-invariant definitions for any metric or
matter perturbations for comoving observers on uniform density
hypersurfaces. As an example we give the curvature perturbation. To
first-order, using Eqs.~(\ref{defpsi_1}) and~(\ref{xi0rho}),
we recover the well-known expression \cite{Bardeen88,WMLL}
\be
\label{defzeta1}
- \zeta_1 \equiv
\widetilde{\psi_1}\Big|_{\rho}
=\psi_1+\H\frac{\drho}{\rho_0'}\,.
\ee
For the second order curvature perturbation on uniform density
hypersurfaces, along comoving worldlines, we use Eq.~(\ref{defpsi_2})
with Eqs.~(\ref{xi0rho}), (\ref{beta0rho}) and (\ref{betai}), to give
\bea
\label{defzeta2}
- \zeta_2 &\equiv& \widetilde{\psi_2}\Big|_{\rho} \nonumber\\
&=&\psi_2+\frac{\H}{\rho_0'}\drhorho 
-2\frac{\H}{{\rho_0'}^2}\drho'\drho
-2\frac{\drho}{\rho_0'}\left(\psi_1'+2\H\psi_1\right)\nonumber\\
&&+\frac{{\drho}^2}{{\rho_0'}^2}\left(\H\frac{\rho_0''}{\rho_0'}
-\H'-2\H^2\right) 
 \nonumber\\
&&+2\left(\psi_1+\H\frac{\drho}{\rho_0'}\right)_{,i}\widetilde\beta_1^i
\,.
\eea

We can also give gauge-invariant definitions for scalar quantities on
uniform density hypersurfaces with comoving worldlines. As an example
we write the
pressure perturbation on uniform-density hypersurfaces
at first order
\be
\label{deltaPnad1}
\widetilde{\dP} \Big|_{\rho} =
 \dP - \frac{P_0'}{\rho_0'} \drho \,.
\ee
%
We can identify this as the usual gauge-invariant definition of the 
non-adiabatic part of the pressure perturbation. 
At second order we have
%
%
%
\bea
\label{deltaPnad2}
\widetilde{\dPP}\Big|_{\rho} 
&=&
\dPP-\frac{P_0'}{\rho_0'}\drhorho 
 + 2\left( \dP-\frac{P_0'}{\rho_0'}\drho \right)_{,i}
 \widetilde\beta_1^i \nonumber\\
&&+P_0'\left\{
\left(\frac{\drho'}{\rho_0'}-\frac{\dP'}{P_0'}\right)
\frac{2\drho}{\rho_0'} 
+\left(\frac{P_0''}{P_0'}-\frac{\rho_0''}{\rho_0'}\right)
\frac{{\drho}^2}{\rho_0'^2}\right\}  \,.\nonumber \\
\eea

These results are readily extended to systems involving scalar fields,
as scalar fields obey the same transformation rules as the energy
density, given in Eqs.~(\ref{defrho1}) and (\ref{defrho2}). Hence one
can write down the comoving curvature perturbation (i.e., the
curvature perturbation on uniform scalar field hypersurfaces) or the
relative entropy perturbation between two fields.

For adiabatic perturbations the local pressure is a unique function of
the local density, $P=P(\rho)$. Hence we can identify the
non-adiabatic part of the pressure perturbation, to first and second
orders, as
\bea
\dP_{\rm nad} &=& \dP - c_s^2 \drho \,, \nonumber \\
\dPP_{\rm nad} &=& \dPP - c_s^2 \drhorho -
 \frac{dc_s^2}{d\rho_0} \drho^2 \,, 
\eea
where the extra term on the right-hand-side of the expression for the
second-order non-adiabatic pressure perturbation
arises from the local variation of the adiabatic sound speed,
$c_s^2\equiv dP/d\rho$.
Note that the non-adiabatic pressure perturbation is automatically
gauge-invariant at first order, but the second order non-adiabatic
perturbation is only gauge-invariant if the first order non-adiabatic
pressure perturbation vanishes~\cite{marco}.

Thus we can write the gauge-invariant pressure perturbation on uniform
density hypersurfaces as
\bea
\label{defP1}
\widetilde\dP|_\rho &=& \dP_{\rm nad} \,,\\
\label{defP2}
\widetilde\dPP|_\rho &=& \dPP_{\rm nad}
 - 2\frac{\drho}{\rho_0'}\dP_{\rm nad}' \,.
\eea
We see that the pressure perturbation will vanish on hypersurfaces of
uniform density for adiabatic perturbations.

\subsubsection{Uniform curvature hypersurfaces}

Instead of defining quantities on uniform density hypersurfaces we can
choose to work with uniform curvature slices. In some scenarios this
can have the advantage of staying non-singular even when the uniform
density hypersurfaces become ill-defined \cite{brandenberger,MWU}.

We can define uniform curvature hypersurfaces by $\widetilde{\psi_1}=0$
and $\widetilde{\psi_2}=0$, which fixes the temporal gauge shift and
$\widetilde{E}_1=0$ and $\widetilde{F}_1^i=0$ to fix the spatial gauge shift
to first-order. 
This implies for the first order temporal gauge
shift, using Eq.~(\ref{defpsi_1}),
\be
\label{xi0psi}
\bar\alpha_1=\frac{\psi_1}{\H}\,,
\ee
and at second order, using Eq.~(\ref{defpsi_2}),
\be
\label{xi0_2psi}
\bar\alpha_2
=
\frac{1}{\H}\left[
\psi_2+2\psi_1^2+\frac{1}{\H}\psi_1'\psi_1+\psi_{1,k}\bar\beta_1^k
\right]\,,
\ee
%
%
where we used Eq.~(\ref{xi0psi}) and fix the spatial gauge
shift~\cite{KS,thesis} 
\be
\bar\beta_1^i = E_{1,}^{~~i} + F_1^i \,.
\ee

The density perturbation on uniform curvature
hypersurfaces is then, at first order, using Eqs.~(\ref{defrho1}),
\be
\widetilde{\drho}\Big|_{\psi}
\equiv\drho+\frac{\rho_0'}{\H}\psi_1\,.
\ee
%
%
At first order the curvature perturbation defined on uniform density
hypersurfaces and the density perturbation on uniform curvature
hypersurfaces are simply related by
\be 
\widetilde{\drho}\Big|_{\psi}
=\frac{\rho_0'}{\H}\widetilde{\psi_1}\Big|_{\rho}\,.
\ee

{}From Eq.~(\ref{defrho2}) we get the definition
of the second order density perturbation on uniform curvature
hypersurfaces
\bea
\label{defrho2_psi}
\widetilde{\drhorho}\Big|_{\psi}
&\equiv&\drhorho+\frac{\rho_0'}{\H}\psi_2
+\frac{1}{\H}\left(
2\rho_0'+\frac{\rho_0''}{\H}-\frac{\rho_0'\H'}{\H^2}
\right)\psi_1^2\nonumber \\
&&+2\frac{\rho_0'}{\H^2}\psi_1'\psi_1
+\frac{2}{\H}\psi_1\drho'  \nonumber\\
&&+2\left(
\drho+\frac{\rho'}{\H}\psi_1 \right)_{,i}\bar\beta_1^i \,.
\eea
Again, these results are readily extended to scalar fields which obey
the same gauge transformations rules as the energy density.

\section{Evolution of curvature on large scales}

Having defined gauge-invariant second-order variables, we now turn to
finding evolution equations for these quantities. 

A fundamental question in cosmology is how density perturbations
evolve in the large-scale regime where gravitational 
perturbations
cannot be neglected. In particular it is important to establish
whether non-linear evolution could introduce significant
non-Gaussianity, e.g., in CMB anisotropies, even in inflationary
models where the large scale structure of the universe is supposed to
arise from purely Gaussian fluctuations at very early times.

Despite the complexity of the field equations at second order (see
e.g.~Ref.~\cite{Noh}) it is sufficient to use the local conservation
of energy-momentum to establish the conservation of $\zeta$ if we
neglect spatial gradients, which we expect to be valid on sufficiently
large scales. This strategy was first employed in Ref.~\cite{WMLL} to
establish the conservation of $\zeta$ at first order.
We will also neglect terms quadratic in first-order vector and tensor
perturbations which can be calculated using the first-order equations
of motion and can be shown to be small on large scales in an expanding
universe. We defer a full treatment including the effect of
these quadratic terms to future work.

%

We define the energy momentum tensor as
\be
\label{defTmunu}
T^{\mu\nu}\equiv\left(\rho+P\right)u^\mu u^\nu+P g^{\mu\nu}
+\Pi^{\mu\nu}\,,
\ee
where $\Pi^{\mu\nu}$ is the trace-free anisotropic stress tensor,
and $u^\mu$ is the fluid 4-velocity.

Energy conservation is given by
%
$u_\nu\nabla_\mu T^{\mu\nu}=0$.
In the homogeneous background we have
\be
 \rho_0'+3\frac{a'}{a}\left(\rho_0+P_0\right)=0 \,,
\ee
while dropping spatial gradient terms we obtain
%
\bea
\drho'+3\H\left(\drho+\delta P_1\right)
-3\left(\rho_0+P_0\right)\psi_1'&\simeq&0 \,, \\
\drhorho'+3\H\left(\drhorho+\delta P_2\right)
-3\left(\rho_0+P_0\right)\psi_2' &&\nonumber \\
\qquad-6\psi_1'\left[\drho+\delta P_1
+2\left(\rho_0+P_0\right)\psi_1\right]
&\simeq&0 \,,
\eea
to first and second order respectively. These can be written in terms
of the gauge-invariant curvature perturbation, $\zeta$ defined in
Eqs.~(\ref{defzeta1}) and~(\ref{defzeta2}), giving
\be
\zeta_1' \simeq - \frac{\H}{(\rho+P)} \widetilde\dP|_\rho
\,,
\ee
and
\bea
\zeta_2' &\simeq& - \frac{\H}{(\rho+P)} \widetilde\dPP|_\rho \nonumber \\
&&- \frac{2}{\rho_0+P_0}
 \left[ \widetilde\dP|_\rho - 2(\rho_0+P_0)\zeta_1 \right] \zeta_1' \,,
\eea
where the gauge-invariant pressure perturbation on uniform density
hypersurfaces is given in terms of the non-adiabatic pressure
perturbation in Eqs.~(\ref{defP1}) and~(\ref{defP2}).
Thus the curvature perturbation on uniform density hypersurfaces is
constant at first and second order for adiabatic perturbations on
large scales if we can neglect spatial gradients~\footnote{
An alternative argument for the existence of a conserved quantity at
second order was given recently in Ref.~\cite{LW03} defined in terms
of the density perturbation on uniform expansion hypersurfaces. One
can verify that for adiabatic perturbations the second-order part
of the conserved quantity defined in Eq.~(11) of Ref.~\cite{LW03}
coincides with $\zeta_2-2\zeta_1^2$, using the gauge-invariant
defintions of $\zeta_1$ and $\zeta_2$ given in
Eqs.~(\ref{defzeta1}) and~(\ref{defzeta2}), written in terms of the
density perturbation evaluated on spatially flat hypersurfaces.}.

Hence we expect an initially Gaussian distribution of adiabatic
curvature perturbations in the very early universe will remain
Gaussian in the large scale limit. By the same token, primordial
non-Gaussianity of the perturbations may be indicative of
non-adiabatic evolution in the early universe.

\section{Conclusion}

In summary, we have given a procedure for defining gauge-invariant
cosmological perturbations at first and second order. As an example we
have given a gauge-invariant definition of the curvature perturbation on
uniform density hypersurfaces. We expect that this prescription could
be extended to higher orders if desired.

We were then able to show, using only the local energy conservation
equation, that the curvature perturbation remains constant on large
scales for adiabatic perturbations where we neglect spatial
gradients.
As shown in Ref.~\cite{LW03}, a conserved perturbation exists of any
quantity that obeys an autonomous local conservation equation. Thus we
can construct conserved perturbations of the local energy density of
any fluid with a barotropic equation of state.

{\em Note added:} While writing up this letter, a non-linear result
for the constancy of $\zeta$ in a long-wavelength
approximation~\cite{Salopek} was
reported by Rigopoulos and Shellard~\cite{RS}.

\acknowledgments

The authors are grateful to 
Marco Bruni, Kouji Nakamura, David Lyth,
David Matravers and Toni Riotto for useful discussions.  This work was
supported by PPARC grant \emph{PPA/G/S/2000/00115}.  KM is supported
by a Marie Curie Fellowship under the contract number
\emph{HPMF-CT-2000-00981}. DW is supported by the Royal Society.
Algebraic computations of tensor components were performed using the
GRTensorII package for Maple.

{}

\end{document}